\documentclass[conference]{IEEEtran}
\IEEEoverridecommandlockouts
\usepackage{cite}
\usepackage{amsmath,amssymb,amsfonts}
\usepackage{algorithm}
\usepackage{algorithmic}
\usepackage{graphicx}
\usepackage{textcomp}
\usepackage{xcolor}
\def\BibTeX{{\rm B\kern-.05em{\sc i\kern-.025em b}\kern-.08em
    T\kern-.1667em\lower.7ex\hbox{E}\kern-.125emX}}

\begin{document}

\title{Multicore Quantum Architecture for Scalable and Energy-Aware Green IoT Computing}

%\title{Suggestion: Communication-Aware Classical and Quantum Multi-Core Architectures for Green IoT}

\author{\IEEEauthorblockN{1\textsuperscript{st} Seham Al Abdul Wahid}
\IEEEauthorblockA{\textit{Electrical and Computer} \\
\textit{Engineering Department} \\
\textit{Toronto Metropolitan University}\\
Toronto, Canada \\
salabdulwahid@torontomu.ca}
\and
\IEEEauthorblockN{2\textsuperscript{nd} Farah Mohammadi}
\IEEEauthorblockA{\textit{Electrical and Computer} \\
\textit{Engineering Department}\\
\textit{Toronto Metropolitan University}\\
Toronto, Canada \\
fmohamma@torontomu.ca }
\and
\IEEEauthorblockN{3\textsuperscript{rd} Yazan Otoum}
\IEEEauthorblockA{\textit{School of Computer Science} \\
\textit{and Technology} \\
\textit{Algoma University} \\
Brampton, Canada \\
otoum@algomau.ca }
\and
\IEEEauthorblockN{4\textsuperscript{th} Arghavan Asad}
\IEEEauthorblockA{\textit{School of Computer Science} \\
\textit{and Technology} \\
\textit{Algoma University} \\
Brampton, Canada \\
arghavan.asad@algomau.ca }
}

\author{
\IEEEauthorblockN{
Seham Al Abdul Wahid\IEEEauthorrefmark{1},
Farah Mohammadi\IEEEauthorrefmark{1},
Yazan Otoum\IEEEauthorrefmark{4},
and Arghavan Asad\IEEEauthorrefmark{4}
}

\IEEEauthorblockA{
\IEEEauthorrefmark{1}Electrical and Computer Engineering Department, Toronto Metropolitan University, Toronto, Canada
}

\IEEEauthorblockA{
\IEEEauthorrefmark{4}School of Computer Science and Technology, Algoma University, Brampton, Canada
}
}

\maketitle

\begin{abstract}
The rapid growth of the Internet of Things (IoT) is increasing the computational demand placed on edge and backend processing infrastructures, motivating scalable architectures that can reduce execution and communication overhead. This paper presents a communication-aware comparison of homogeneous multicore classical and quantum architectures for Green IoT using the Qcomm simulator. In addition, a mesh-based, homogeneous, quantum-classical multicore architecture is proposed to enable scalable, efficient execution of distributed quantum workloads. We evaluate a 4×4 mesh of 16 processing cores under regular, classical-bottleneck, and quantum-bottleneck conditions, and study the effects of traffic pattern and Light-to-Matter (LTM) port provisioning on performance. Results show that butterfly traffic yields the largest gains over uniform traffic in both execution time and qubit coherence, while performance benefits from additional LTM ports saturate beyond four ports. At the algorithm level, homogeneous Multi Quantum Processing Units (QPUs) achieve $10\times$, $9\times$, and $7\times$ energy efficiency improvements in comparison to Multi-CPUs for Quantum Approximate Optimization Algorithm (QAOA), Variational Quantum Eigensolver (VQE), and Quantum Neural Network (QNN), respectively. The paper's findings show that communication-aware multicore quantum architectures are promising for latency-sensitive IoT processing. Furthermore, the results validate the proposed architecture and provide a foundation for future scheduling and resource management techniques to improve scalability and efficiency in multi-QPU systems. 
\end{abstract}


%\textcolor{red}{Yazan: 1- The title and overall framing suggest an energy-aware or Green IoT comparison; however, the results only report execution time. 2- The abstract is short and so general. Please see the suggested updated abstract below. 3- Please list the contribution of this paper at the end of the Introduction section. 4- Add the paper structure at the end of the introduction section. 5- The bullet definitions of quantum sensors, repeaters, gateways, and encryption read more like a survey than a technical paper, and I don't feel it is necessary. We need to save the space for more technical depth, such as the system model. 6- The Methods section is weak and short, and it needs to be extended with more details, such as adding the exact parameter values. answering some questions like: How were QAOA, VQE, and QNN circuits mapped onto the 16-core architecture? How many runs were averaged? And are the results deterministic or averaged over seeds? 7- The quality of the figures is low in general, and they need to be enhanced, as they are not readable sometimes. 8- After we polish it, we need to revise the conclusion. 9- The paper length should be 6 pages. The current version is short for the GC. 10- The self-citation is currently high. Consider adding more state-of-the-art references. }

\vspace{-2mm}
\section{Introduction}
The Internet of Things (IoT) marks a new era of connectivity, enabling real-time decision-making and enhancing the quality of life across domains such as smart homes, intelligent transportation, smart cities, and healthcare. However, due to the widespread use of IoT devices and their wireless communication characteristics, they are exposed to internal and external security threats \cite{11146534}. In addition, IoT requires real-time data analysis to provide important insights, and implementing it with conventional Machine Learning (ML) and data mining algorithms is challenging when massive data collection and analysis are required \cite{https://doi.org/10.1049/qtc2.12079}. The continued use of IoT has also led to the rapid increase in waste, greenhouse gas emissions and the consumption of natural and non-renewable resources. As IoT is expected to further increase the number of network nodes, it is essential to address these challenges and minimize the resources and energy required to implement network elements \cite{ABOUBAKAR20224163}. As a result, the Green Internet of Things (Green IoT) emerged to focus on optimizing and minimizing energy use in IoT networks and reducing their carbon footprint \cite{https://doi.org/10.1049/ntw2.70003}. Quantum computing has the potential to enhance the decision-making capabilities of IoT devices, help prevent cyberattacks, and reduce execution time, thereby reducing energy consumption \cite{adere2026systematic}, \cite{11242178}. A study demonstrated that \textit{n} Quantum Bits (qubits) can encode more information than \textit{n} classical bits, offering the potential for more efficient computation than current methods. However, because current quantum hardware is limited by the number of available qubits, not every node in an IoT network can be replaced with a quantum device. Heterogeneous approaches can balance the trade-off between data size, efficiency, and security \cite{https://doi.org/10.1049/qtc2.12079}, \cite{11206930}. Quantum technologies, coupled with classical devices, can enable faster computation and increase the reliability of IoT networks \cite{russo2025telesabre}.

Current quantum hardware remains in the Noisy Intermediate-Scale Quantum (NISQ) era, characterized by limited qubit counts (tens to hundreds), high error rates, and the absence of full-scale quantum error correction (QEC). As IoT devices require large-scale deployment to handle high data volumes, multicore architectures are necessary. Architectural designs need to enable quantum communication physically between quantum cores. However, as classical information is typically exchanged between communicating quantum cores, recent studies model a Network-on-Chip (NoC) to support teleportation and handle the transmission of control and data. It is important to analyze the impact of this classical communication, which aids inter-core operations, on execution time and overall system efficiency for quantum workload execution \cite{russo2025telesabre}. The primary focus of existing quantum computing simulators, such as Qiskit Aer, is the execution of quantum circuits on a single monolithic processor \cite{10785069}. These tools are effective for algorithm validation and small-scale quantum systems. However, their drawback is the lack of support for modelling distributed architectures in which multiple quantum processing units communicate via quantum or classical channels. This makes the evaluation of architectural tradeoffs challenging. Qcomm is a recently developed simulator that addresses these drawbacks, providing an environment for evaluating the architecture and algorithms at both the circuit and network levels. Qcomm will be used to present the results of this paper, providing valuable insight into a scalable multicore quantum computing architecture for IoT applications \cite{10.1145/3779440}. While significant progress has been made in isolated components of Quantum IoT (QIoT), most existing work focuses on single-core quantum processors or on theoretical advantages, without addressing architectural scalability. In contrast, classical IoT platforms employ multicore architectures to improve throughput and energy efficiency. However, the comparison between classical multicore processors and emerging multicore quantum architectures remains unexplored, especially from a Green IoT perspective. This paper addresses this gap by providing a comparative analysis of homogeneous multicore classical architectures and multicore quantum architectures. This work aims to clarify the role of multicore quantum computing as a complementary and sustainable computing paradigm for future IoT systems. In addition, this paper proposes a threshold-guided multi-QPU architecture that integrates classical control to enable scalable execution of quantum workloads, with a classical fallback when the required qubit resources exceed the available quantum capacity. Accordingly, this study examines Green IoT at the computer architecture level, focusing on how processing, inter-core communication, and resource provisioning influence the energy efficiency of future quantum-enabled IoT backend and edge compute systems. The key contributions of this work are summarized as follows:
\begin{itemize}
\item A comparative analysis of homogeneous multicore classical (Multi-CPU) and multicore quantum (Multi-QPU) architectures.
\item Implementation and evaluation of QAOA, VQE, and QNN workloads on a 4×4 mesh-based architecture using the Qcomm simulator, capturing both circuit-level and network-level behaviour. QAOA, VQE, and QNN are treated as representative computational kernels that may support future IoT optimization, learning, and data-analysis tasks
\item Investigation of the impact of traffic patterns and bottleneck conditions on execution time and qubit coherence.
\item Proposal of a mesh-based homogeneous multicore quantum-classical architecture to enable scalable and communication-efficient quantum workload execution.
\end{itemize}

The remainder of this paper is organized as follows. Section II reviews related work in quantum IoT and multicore architectures. Section III presents the background on quantum IoT and multicore quantum systems. Section IV describes the system model, experimental methodology, and the proposed multicore quantum-classical architecture. Section V presents the results and evaluation under various traffic and bottleneck conditions. Finally, Section VI concludes the paper and outlines directions for future work.

\section{Related Work}
Recent studies have demonstrated the capabilities of quantum computing-based techniques to overcome the current limitations in classical implementations of IoT systems, specifically in the context of scalability, energy efficiency and security. For instance, classical ML algorithms deployed on multicore CPUs and GPUs are often unsuitable for handling the large-scale data required for 6G networks due to their computational and time complexity. Whereas, quantum machine learning techniques can provide an exponential speed-up in the learning process due to their superposition and parallelism properties. QIoT can also incorporate security measures based on Quantum Key Distribution (QKD) algorithms to secure communication between classical and quantum entities \cite{10806885}. QKD is based on the principle of quantum entanglement and the no-cloning theorem, ensuring that data exchanged between devices remains encrypted. A recent study proposed a QIoT-based framework for secure medical data exchange, demonstrating improved data integrity and reduced storage requirements by exploiting quantum information encoding \cite{10631683}. Additional studies focus on modular and multicore quantum architectures, including distributed quantum processors, to improve scalability and fault tolerance. The studies have reported measurable gains in circuit fidelity and feasibility of distributed quantum operations under realistic noise constraints \cite{Main_Drmota_Nadlinger_Ainley_Agrawal_Nichol_Srinivas_Araneda_Lucas_2025}. Recent studies on Green IoT focus on improving the energy efficiency and scalability of classical multicore architectures through techniques such as edge intelligence optimization and communication-efficient learning. These approaches have demonstrated energy reductions of 20-40\% and improved latency for IoT workloads by dynamically adapting core utilization and reducing communication overhead \cite{1a35d294a9924fc4b02f8a2e00a589ee}. However, direct comparisons between classical and quantum multicore architectures using common Green IoT metrics such as energy efficiency, latency, and workload sustainability remain limited, motivating the comparative analysis presented in this work. 

\vspace{-1mm}

\section{Background}
\subsection{Quantum IoT}
Quantum IoT integrates quantum principles into the traditional IoT framework to offer the quantum advantage. It comprises several components, including quantum sensors, quantum communication, quantum processors, quantum repeaters, quantum gateways, and quantum encryption, forming a cohesive system where quantum computation and communication interact with classical control mechanisms to enable efficient workload execution. The combination of these components introduces a new paradigm for secure communication, computation, and sensing that can be implemented across various applications and provides a new definition of green IoT \cite{10631683}, \cite{hassan2026quantum}. In the unified system architecture, quantum sensors enable high-precision data acquisition, while quantum processors execute computationally intensive tasks that are impractical for classical systems. Quantum communication mechanisms, supported by repeaters and encryption techniques, provide secure and reliable data transfer across distributed nodes, and quantum gateways facilitate seamless integration between classical and quantum domains \cite{alipour2025qiot}.

\subsection{Multicore Quantum Architecture}
A multicore architecture allows different tasks to run on different processors simultaneously. The topology is defined by the placement of multiple cores in a specific configuration in which the cores communicate with each other. This paper focuses specifically on the Quantum Network-on-Chip (qNoC), which derives its name from communication among various cores, thereby reducing execution speed. Another method to improve the scalability of quantum computing is a quantum farm, which evenly distributes tasks across multiple quantum computers. However, NoC remains the most attractive option, since quantum farms require qubit information to travel a larger physical distance, increasing the risk of noise and decoherence \cite{CALEFFI2024110672}.

\vspace{-5mm}

\begin{table}[htbp]
\caption{Physical Parameters, Architectural and Micro-Architectural Parameters, and Circuit Properties Used for the Experiments}
\begin{center}
\begin{tabular}{|l|l|}
\hline
\textbf{Parameters} & \textbf{Configuration} \\
\hline
Network topology & mesh \\
\hline
Mesh size & $4 \times 4$ \\
\hline
NoC clock frequency & from 10 MHz to 1 GHz \\
\hline
RAM bandwidth & 128 Gbps \\
\hline
Physical qubits & from 10 to 10{,}000 \\
\hline
Link width & from 2 to 10 bits \\
\hline
Circuit size -- logical qubits & from 10 to 1000 \\
\hline
LTM ports per QC & from 1 to 5 \\
\hline
\end{tabular}
\label{tab:experiment_parameters}
\end{center}
\end{table}
\vspace{-7mm}

\begin{figure}[h!]
    \centering
    \includegraphics[width=0.4\textwidth]{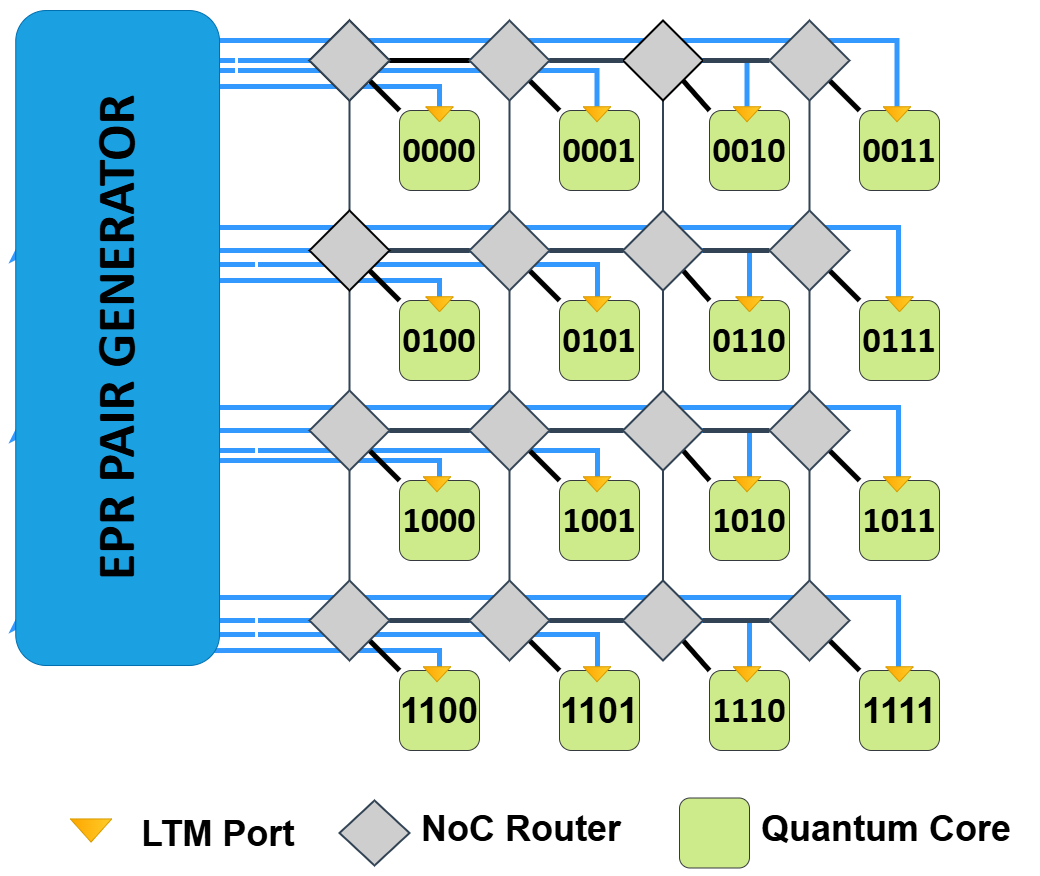}
    \vspace{-2mm}
    \caption{4x4 mesh architecture employed by QCOMM with NoC routers, LTM ports and an EPR pair generator. Each quantum core is connected via a NoC router, while the LTM supports entanglement generation and inter-core quantum communication \cite{10.1145/3779440}.}
\end{figure}

\vspace{-5mm}

\section{Methods}
This paper utilizes Qcomm, a simulation tool, to evaluate the impact of communication under various traffic workload conditions and to investigate three algorithms, Quantum Approximate Optimization Algorithm (QAOA), Variational Quantum Eigensolver (VQE), and Quantum Neural Network (QNN), on Multi-CPUs and Multi-QPUs (quantum processing units). \cite{10.1145/3779440}. QAOA circuits were configured with a depth of 2; VQE circuits had a depth of 3; and QNN models consisted of 3 variational layers with parametrized rotation gates and entanglement blocks. Each circuit was constructed using a combination of single-qubit rotation gates and two-qubit entangling gates, representative of NISQ-era workloads. Table I summarizes Qcomm's simulator architecture for multi-QPUs and multi-CPUs. The circuit sizes ranged from 10 to 1000 logical qubits and were mapped onto the available physical qubits per core, as defined in Table I, ensuring feasibility within the simulated architecture. The larger qubit configurations in Table I are simulation-scale projections used to study architectural scalability and do not represent currently available multicore quantum hardware. The simulator is set up in a 4 × 4 mesh configuration, as shown in Fig. 1. This means the network comprises 16 cores, and each quantum core is assigned 1 NoC router to transfer classical data between cores. The mapping of quantum circuits onto the 16-core architecture was performed by distributing logical qubits across cores according to locality and connectivity constraints, while inter-core operations were handled via teleportation, supported by EPR pair generation through the Light-to-Matter (LTM) ports and NoC routing. Each core has a certain number of LTM ports, which send qubits to an EPR generator. The EPR generator then entangles two qubits at different cores as needed. This enables distributed execution, in which portions of a quantum circuit run in parallel across multiple cores, while synchronization and communication overhead is captured by the NoC model. Three trials of various synthetic traffic patterns were simulated with 200, 400, 600, 800, and 1000 gates. For each configuration, three independent simulation runs were conducted, and the reported results are then averaged across these runs. To account for stochastic variations in traffic injection and quantum operations, different random seeds were used for each run, yielding non-deterministic yet statistically consistent outcomes. 

\subsection{Energy Modeling}
To evaluate the energy efficiency of multicore classical and quantum architectures in green IoTs, a first-order energy model is introduced to capture the primary contributors to system-level energy consumption during distributed workload execution. The total energy consumption of the system is modelled as the sum of computation and communication energy across all processing units, as shown in (1). Computational energy accounts for CPU or QPU operations, while communication energy accounts for inter-core communication. 
\vspace{-2mm}
\begin{equation}
E_{\text{total}} = E_{\text{comp}} + E_{\text{comm}}
\label{1}
\end{equation}

For classical multicore architectures, energy consumption is dominated by processor activity and memory or communication overhead within the NoC. The total energy consumption is modelled in (2), where P\textsubscript{CPU} is the average power consumption per CPU core, T\textsubscript{CPU} is the execution time, and E\textsubscript{NoC} is the communication energy across the NoC and is further expressed in (2).

\vspace{-4mm}
\begin{equation}
E_{\text{CPU}} = P_{\text{CPU}} \cdot T_{\text{CPU}} + E_{\text{NoC}}
\label{2}
\end{equation}
\vspace{-5mm}
\begin{equation}
E_{\text{NoC}} = N_{\text{msg}} \cdot H \cdot E_{\text{hop}}
\label{3}
\end{equation}

In (3), N\textsubscript{msg} stands for the number of messages exchanged, H is the average hop count, and E\textsubscript{hop} is the energy per hop. Energy consumption in quantum multicore architectures includes contributions from quantum operations (E\textsubscript{gate}), idle coherence maintenance (E\textsubscript{idle}), and communication overhead ($E_{\text{comm}}^{Q}$), as modelled in equations (4), (5), (6), and (7), respectively. Where N\textsubscript{gates} denotes the number of quantum gates, E\textsubscript{g} is the energy per gate operation, P\textsubscript{idle} is the energy consumed while qubits maintain coherence, T\textsubscript{idle} is the time which qubits maintain coherence, N\textsubscript{EPR} is the number of entangled pairs generated, E\textsubscript{EPR} is the energy cost of entanglement generation (via LTM ports), and E\textsubscript{classical} is the classical control and NoC communication energy.  
\vspace{-2mm}
\begin{equation}
E_{\text{QPU}} = E_{\text{gate}} + E_{\text{idle}} + E_{\text{comm}}^{Q}
\label{4}
\end{equation}
\vspace{-3mm}
\begin{equation}
E_{\text{gate}} = N_{\text{gates}} \cdot E_{\text{g}}
\label{5}
\end{equation}
\vspace{-3mm}
\begin{equation}
E_{\text{idle}} = T_{\text{idle}} \cdot P_{\text{idle}}
\end{equation}
\vspace{-3mm}
\begin{equation}
E_{\text{comm}}^{Q} = N_{\text{EPR}} \cdot E_{\text{EPR}} + E_{\text{classical}}
\end{equation}

\subsection{Proposed Architecture}
\vspace{-5mm}
\begin{figure}[h!]
    \centering
    \includegraphics[width=0.35\textwidth]{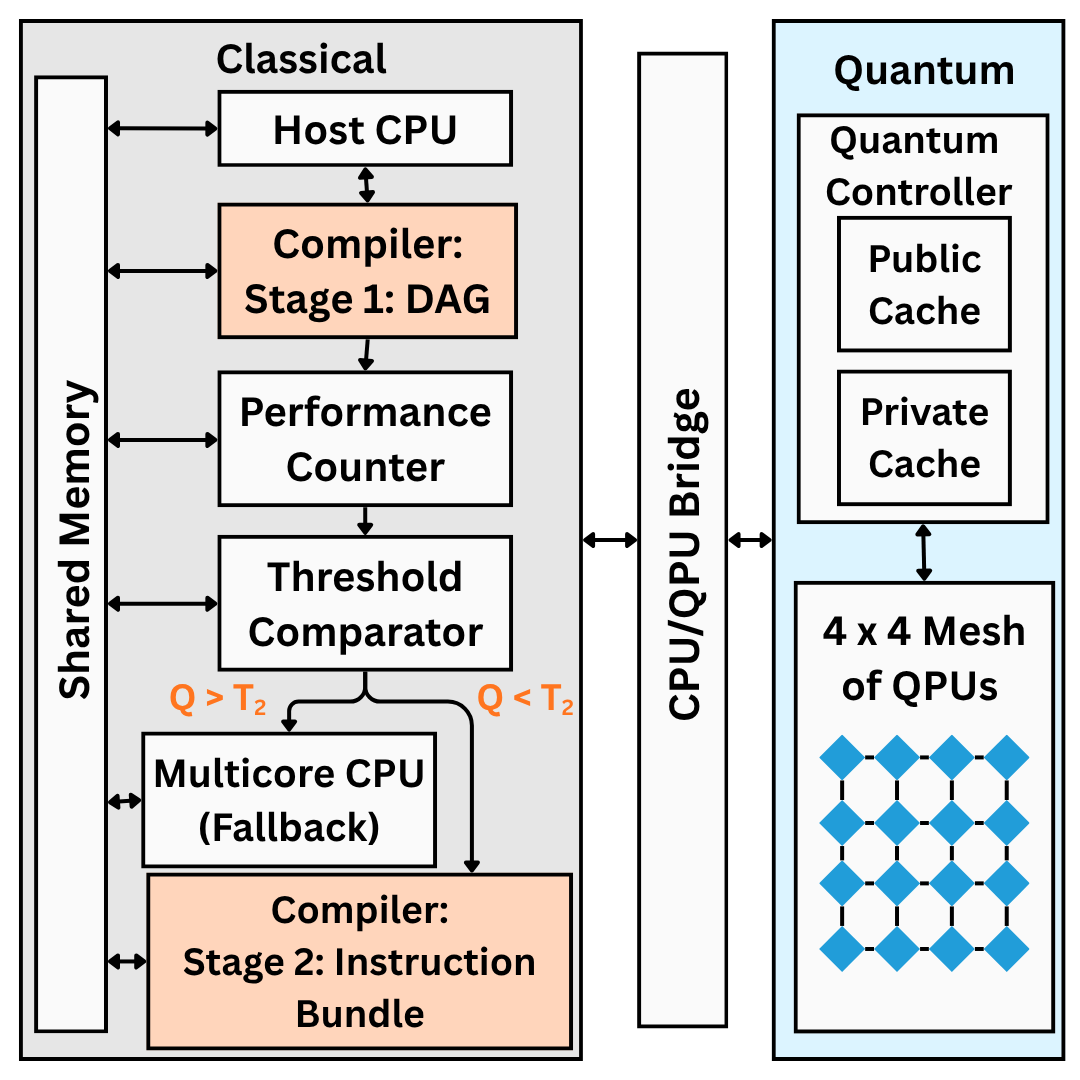} 
    \vspace{-2mm}
    \caption{Proposed Threshold-Based Architecture.}
    \label{Method}
\end{figure}

As multicore quantum architectures are expected to enable scalable, efficient computation for future IoT and high-performance workloads, a mesh-based, multicore quantum-classical architecture is proposed to support distributed execution, minimize communication overhead, improve resource utilization, and maintain a classical fallback path when the quantum workload exceeds the available QPU capacity. The proposed architecture, as shown in Fig. \ref{Method}, consists of the following main components: a compiler, a host CPU, a mesh-based multicore QPU layer, and CPU-QPU communication interfaces. Within a future Green IoT deployment, sensors would perform data acquisition and preprocessing, while computationally demanding, quantum-compatible tasks could be forwarded to the proposed multicore quantum-classical processing layer.

The compiler first translates high-level quantum programs into a hardware executable format through a two-stage process. It is essential that the compiler run on the host side, as it requires access to program memory and architectural parameters before transferring the workload to the quantum chip. In the first stage, the program is converted into a Directed Acyclic Graph (DAG), where nodes represent quantum gates and edges represent data dependencies, enabling instruction-level parallelism, gate reordering, and noise-aware optimization. As illustrated in Algorithm 1, a performance counter is used at this stage to estimate the required quantum resources and qubits before execution and determine the workload. The number of qubits is then compared with two architectural thresholds, $T_1$ and $T_2$. Where $T_1$  equates to the qubit capacity of the target QPU technology, and $T_2$ is 16 times the value of $T_1$, identifying the total available qubits for the proposed 4 x 4 multi-QPU architecture. If the number of required qubits for quantum execution is less than $T_1$, then it is scheduled for a single-core QPU execution. This prevents inter-core communication, reducing synchronization overhead, and is used for low-width IoT workloads. If the number of qubits falls between $T_1$ and $T_2$, the workload is scheduled for execution on a multicore QPU. Highly interacting qubits are placed on the same neighbouring QPUs to reduce long-distance communication. Lastly, if the number of qubits exceeds $T_2$, quantum execution is not possible, and the task is scheduled for classical execution to ensure that the system remains functional.

In the second stage, after the compiler determines the execution mode, the DAG is mapped to hardware-oriented instruction bundles, where each layer corresponds to a set of parallel operations annotated with quantum-core identifiers, gate types, and physical-qubit addresses. The instruction bundle, therefore, provides a hardware-level representation that the quantum controller can decode and dispatch to the compute units. The host CPU acts as the global controller, coordinating the compiler, instruction bundles, synchronization across QPUs, processing measurement results, and activating the classical fallback path when required. In addition, the CPU maintains the program, DAG, instruction queue, and measurement buffers in the shared memory. The shared buffer allows the host CPU, compiler and communication bridge to exchange program metadata, compiled instruction bundles, and execution feedback without requiring direct access to the quantum chip memory. The CPU-QPU communication interface provides the physical bridge between classical and quantum domains. From the classical side, it transfers DAG metadata, compiled instruction bundles, qubit layout information and control signals to the quantum controller. From the quantum side, it returns measurement results, execution status, and feedback signals to the host CPU. This bidirectional interface is essential as the QPU cannot operate independently and requires classical control for scheduling and measurement interpretation.

On the quantum side, the quantum controller receives the instruction bundles and manages their execution on the quantum chip. The controller contains a public cache that stores architecture-visible information, such as the compiled program, qubit layout, and measurement configuration. Additionally, the controller contains a private cache which stores low-level execution information, such as shot count, execution queue entries and quantum error-correction metadata. The compute units inside the controller decode the incoming instruction bundles from the compiler and generate the appropriate control signals for the QPU cores. The quantum chip consists of a 4 × 4 mesh of homogeneous QPU cores, as mentioned in the sections above.

\begin{algorithm}
\caption{Compiler - Scheduling Technique}
\label{alg1}
\begin{algorithmic}[1]
\STATE \textbf{Input:} Quantum program $P$, thresholds $T_1$, $T_2$
\STATE Generate $DAG(P)$ and estimate $q_{\mathrm{req}}$

\IF{$q_{\mathrm{req}} \leq T_1$}
    \STATE Select single-QPU execution
\ELSIF{$q_{\mathrm{req}} \leq T_2$}
    \STATE Select distributed 16-QPU execution
\ELSE
    \STATE Select classical fallback execution
\ENDIF

\STATE Generate and dispatch the corresponding instruction bundles
\end{algorithmic}
\end{algorithm}
 \vspace{-5mm}

\section{Results and Evaluation}
The impact of communication patterns on multicore quantum architecture is evaluated under zero-bottleneck, classical-bottleneck and quantum-bottleneck conditions. Regular denotes the baseline configuration without an intentionally imposed communication bottleneck. Classical bottleneck represents conditions in which NoC communication limits execution, whereas quantum bottleneck represents conditions in which quantum communication processes, including EPR generation and associated inter-core operations, dominate the communication latency. Uniform traffic distributes communication approximately evenly among cores, while bit-reversal maps each source to a destination obtained by reversing its binary node address. Hotspot traffic concentrates communication toward a limited set of cores, whereas butterfly traffic represents a structured permutation pattern that creates deterministic inter-core communication paths. As shown in Fig. 3 and Fig. 4, the butterfly traffic pattern consistently achieves the largest reduction in execution time and the highest improvement in qubit coherence relative to uniform traffic across bottleneck scenarios. This behaviour highlights its suitability for distributed quantum workloads with structured communication. As quantum parameters improve, performance across traffic patterns converges, indicating reduced sensitivity to communication overhead. Notably, hotspot traffic exhibits the greatest relative improvement, transitioning from 6\% worse to 7\% better in execution time and from 17\% worse to 3\% better in qubit coherence compared to bit reversal. This trend demonstrates that localized communication patterns become increasingly advantageous as quantum hardware and entanglement generation improve, while long-distance patterns remain constrained by classical routing effects. The role of parallelization is further examined by varying the number of LTM ports in the NoC architecture. As shown in Fig. 5 and Fig. 6, increasing the number of LTM ports improves performance by enabling concurrent entanglement generation across cores. However, performance gains saturate at approximately four LTM ports, beyond which additional ports provide diminishing returns. The results emphasize that balanced interconnect provisioning is more effective than simply increasing hardware resources for scalable multicore quantum systems. Fig. 7 compares energy cost, using the previously described energy modelling functions, for Multi-CPUs and Multi-QPUs across the QAOA, VQE, and QNN algorithms and their classical counterparts for classical execution. It is evident that Multi-CPUs have a significantly higher energy consumption across all three algorithms. QAOA is approximately 10x more efficient when executed on Multi-QPUs, followed by 9X and 7X improvements for the VQE and QNN algorithms, respectively. These observed trends highlight the potential to enhance Green IoT through Multi-QPU execution, in which circuit execution and measurement can be efficiently distributed across multiple quantum cores.

\begin{figure}[h!]
    \centering
    \includegraphics[width=\columnwidth]{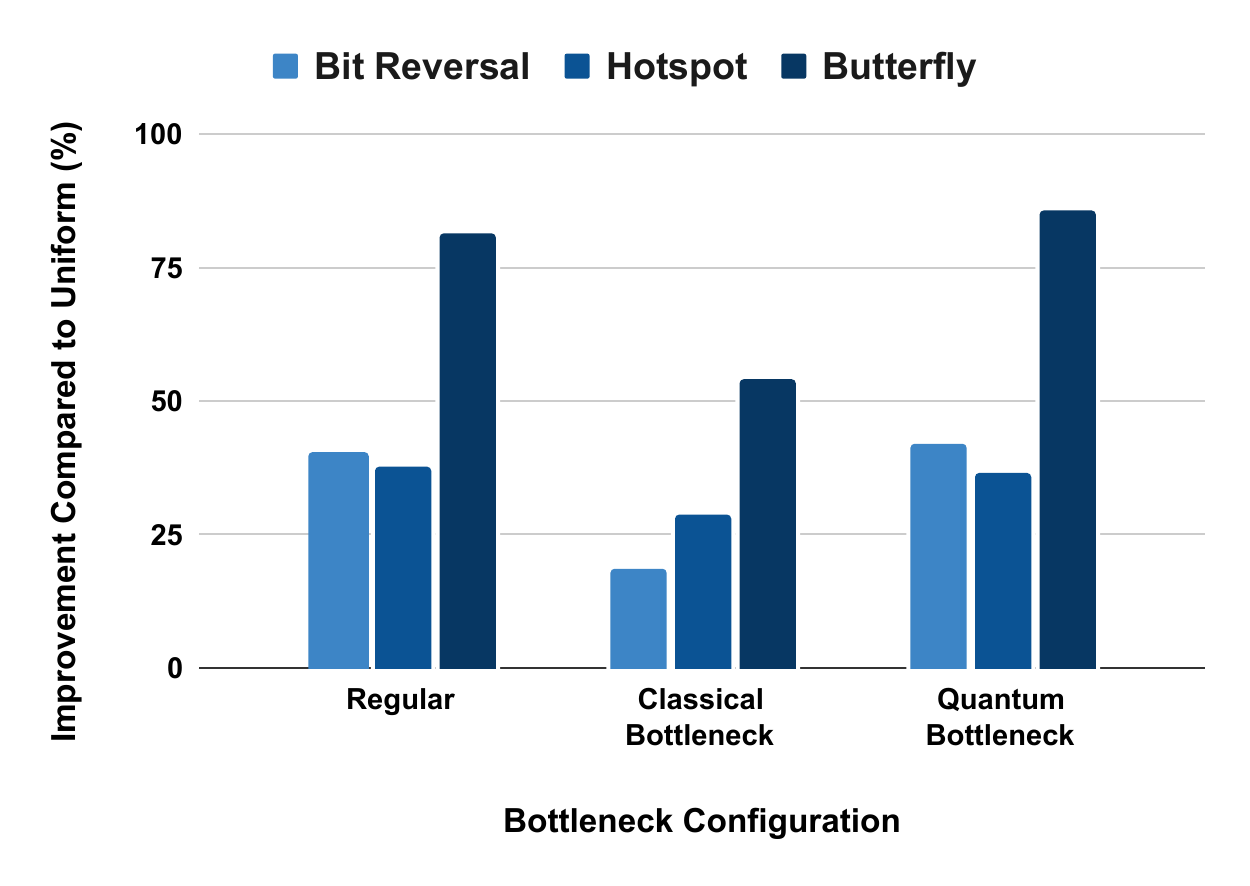}
    \vspace{-10mm}
    \caption{Percentage decrease in execution time of bit reversal, hotspot, and butterfly relative to uniform under zero, classical, and quantum bottleneck conditions.}
\end{figure}
\vspace{-7mm}
\begin{figure}[h!]
    \centering
    \includegraphics[width=\columnwidth]{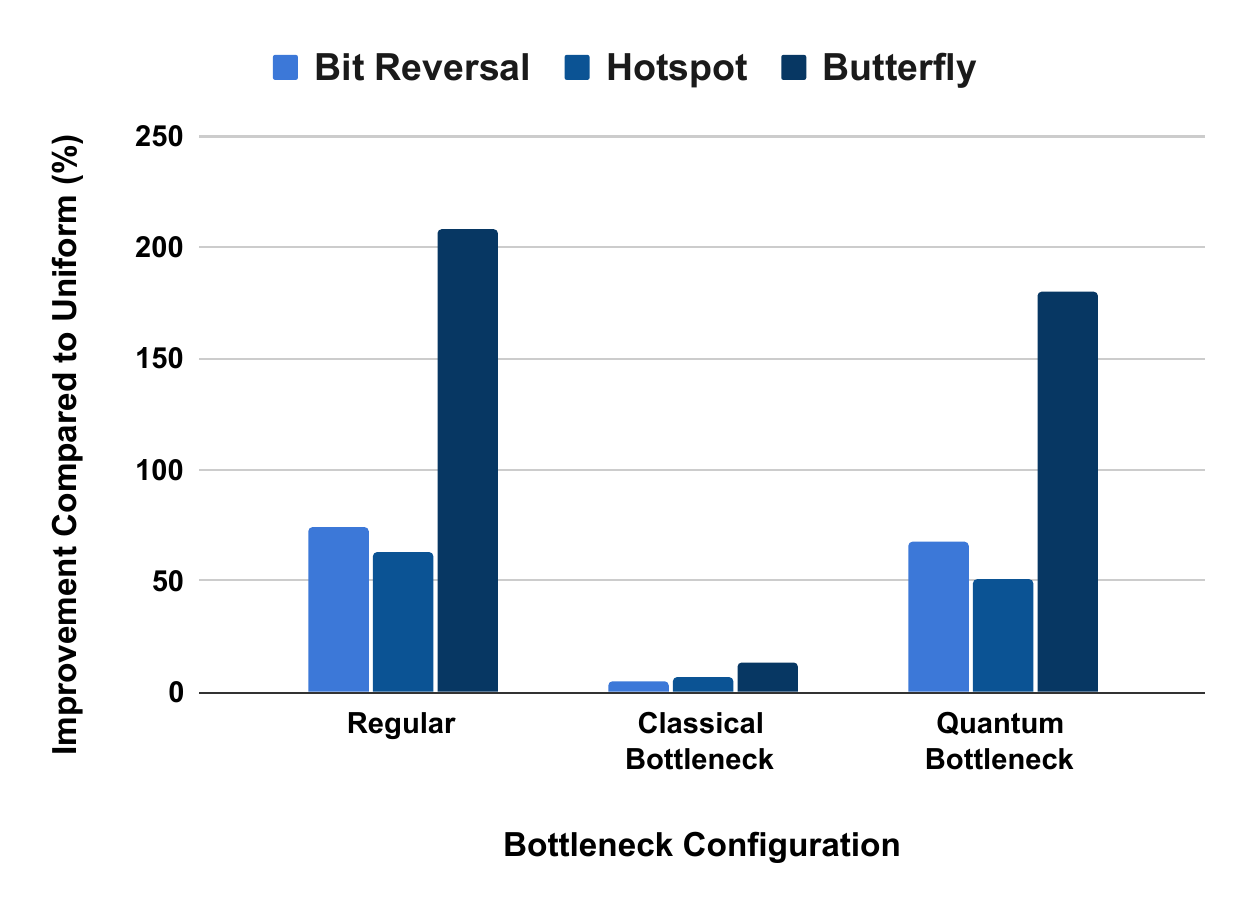}
    \vspace{-10mm}
    \caption{Percentage increase in qubit coherence of bit reversal, hotspot, and butterfly relative to uniform under zero, classical, and quantum bottleneck conditions}
\end{figure}

\begin{figure}[h!]
    \centering
    \includegraphics[width=\columnwidth]{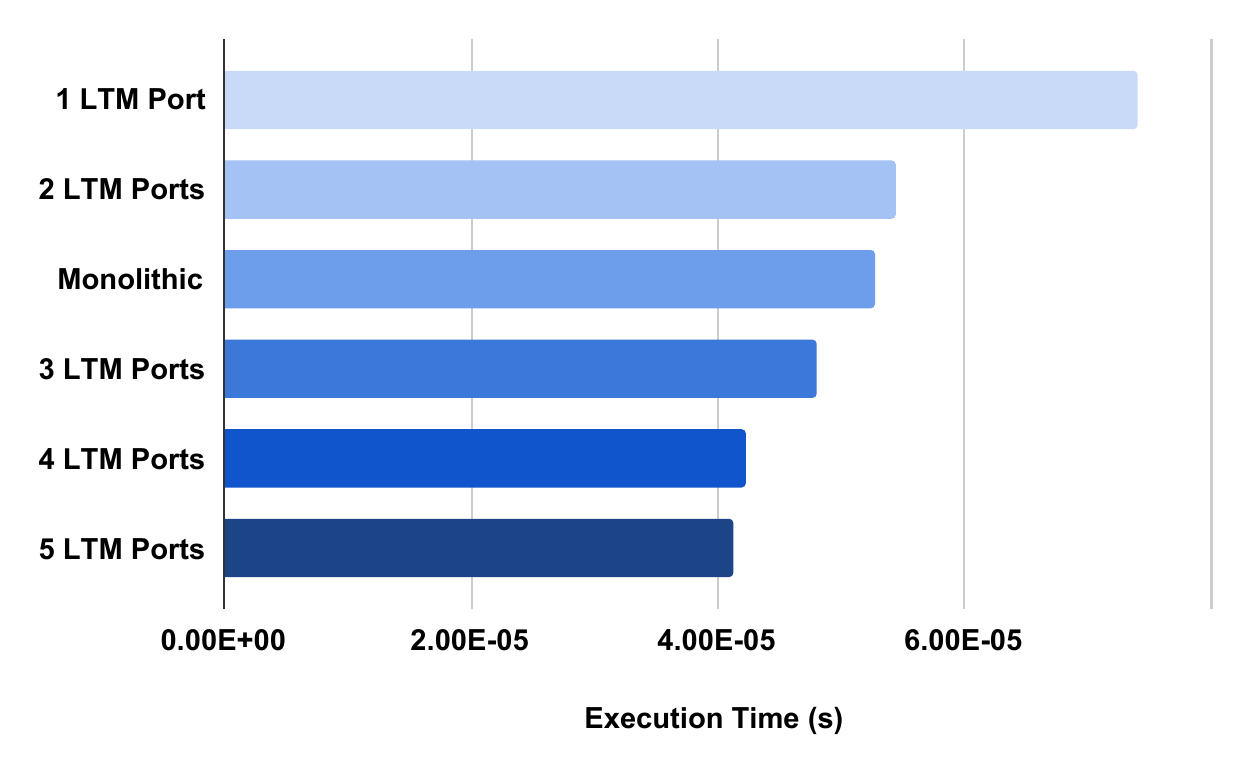}
    \vspace{-10mm}
    \caption{Difference in Execution Time for NoC with Varying LTM Ports and Monolithic Architecture}
\end{figure}
\vspace{-5mm}
\begin{figure}[h!]
    \centering
    \includegraphics[width=\columnwidth]{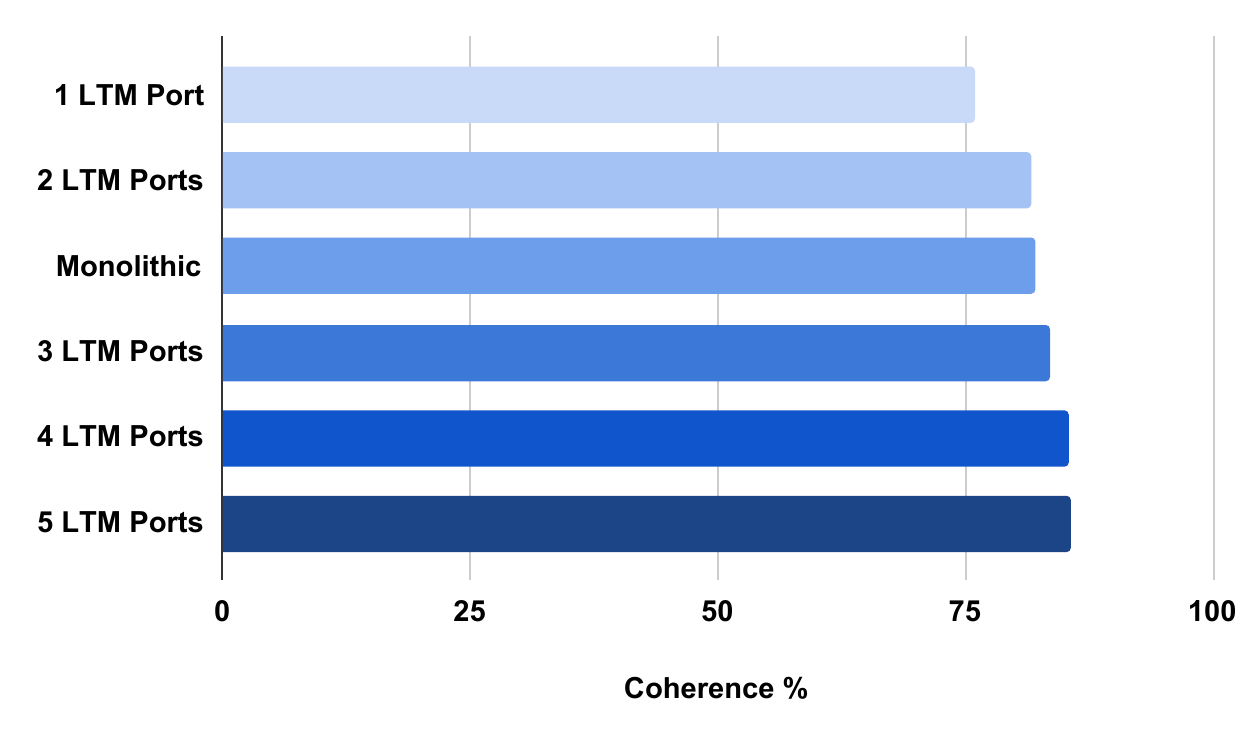}
    \vspace{-10mm}
    \caption{Difference in Coherence for NoC with Varying LTM Ports and Monolithic Architecture}
\end{figure}
\vspace{-5mm}
\begin{figure}[h!]
    \centering
    \includegraphics[width=\columnwidth]{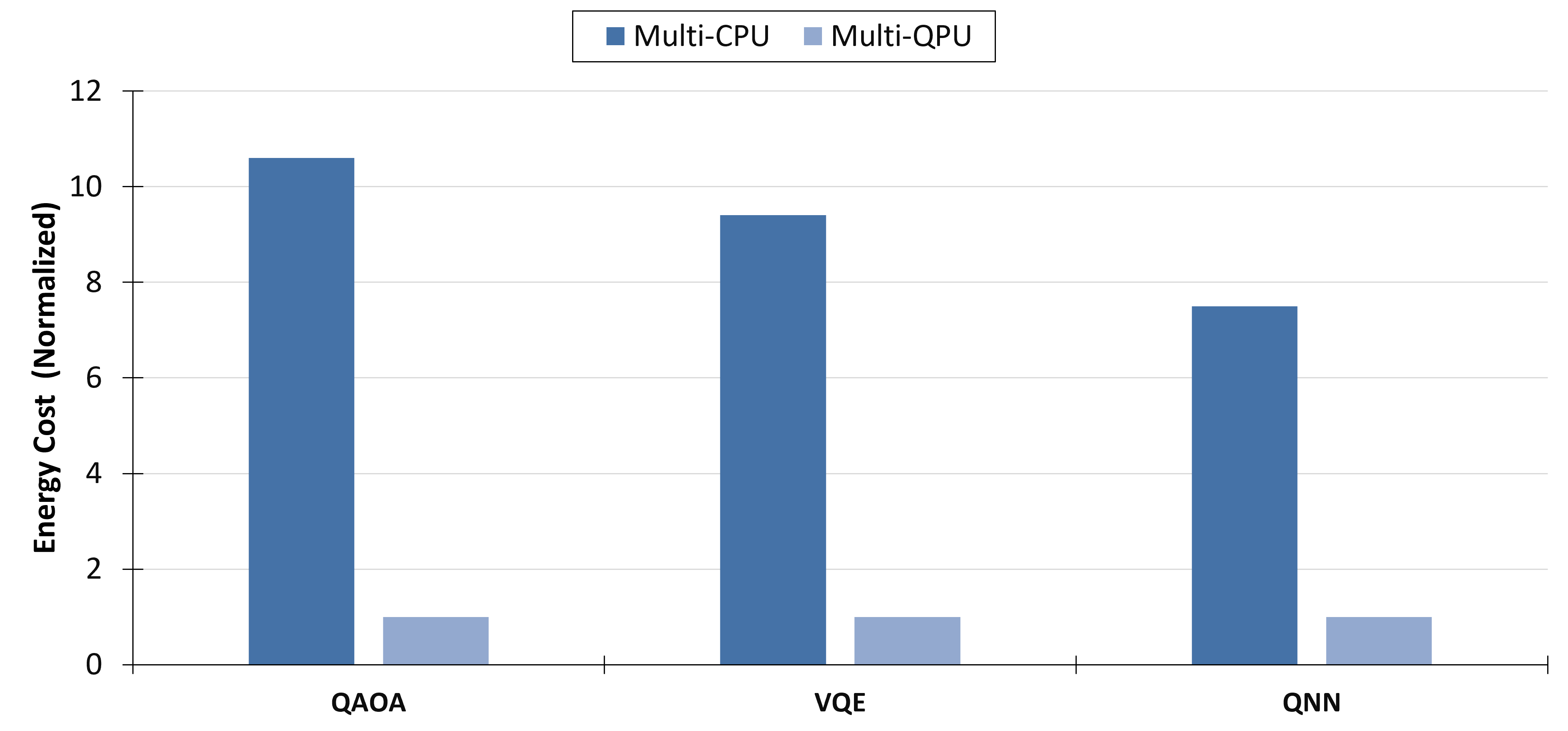} 
    \vspace{-7mm}
    \caption{Comparison of energy consumption in Multi 16 CPUs vs. Multi 16 QPUs for QAOA, VQE and QNN algorithms.}
\end{figure}

\vspace{3mm}
\section{Conclusion and Future Work}
This paper presented a comparative evaluation of homogeneous multicore classical and multicore quantum architectures from a Green IoT perspective. Using the Qcomm simulation framework, the results demonstrate that Multi-QPUs significantly outperform Multi-CPUs in terms of execution time and energy efficiency for quantum algorithms. In addition, the analysis revealed that structured communication patterns and balanced provisioning of quantum resources, such as LTM ports, are critical to achieving scalable performance. Furthermore, the results highlight the importance of communication-aware design, where both workload distribution and inter-core communication significantly influence overall system efficiency. The results identify architectural trends that may support more efficient future IoT computing; however, validation with end-to-end IoT workloads and hardware-calibrated energy models is required before system-level Green IoT benefits can be established. To support these findings, a threshold-based multicore quantum-classical architecture was proposed, providing a scalable framework for distributed quantum workload execution. Future work includes implementing heterogeneous computing architectures that integrate CPUs, GPUs, and QPUs into a unified IoT processing framework. This integration enables more efficient use of both classical and quantum resources. Additionally, benchmarking against real-world IoT applications can help quantify the gap between theoretical speedups and near-term deployable performance. Such extensions will help bridge the gap between simulation-based analysis and practical deployment of quantum-enhanced IoT systems.  
\vspace{-2mm}

\bibliographystyle{IEEEtran}
\bibliography{references}

\end{document}